\documentclass[
reprint,
 amsmath,amssymb,
 aps,showpacs,
]{revtex4-1}

\usepackage[T1]{fontenc} 

\usepackage{graphicx}
\usepackage{bm}
\usepackage{float}
\usepackage{comment}
\usepackage{mathtools}
\usepackage{xcolor}
\usepackage{ulem}
\usepackage{cancel}

\newcommand{\mL}{{\mathcal{L}}}
\newcommand{\mE}{{\mathcal{E}}}
\newcommand{\mB}{{\mathcal{B}}}
\newcommand{\mD}{{\mathcal{D}}}

\newcommand{\pt}{{\partial}}

\begin{document}

\preprint{v2 of arXiv:1911.08714 [hep-ph]}

\title{Electron EM-mass melting in strong fields}

\author{Stefan Evans}
\author{and Johann Rafelski}
\affiliation{Department of Physics, The University of Arizona, Tucson, AZ 85721, USA}

\begin{abstract}
We study the response of the electron mass to an externally applied electrical field. As a consequence of nonlinear electromagnetic (EM) effective action, the mass of a particle diminishes in the presence of an externally applied electric field. We consider modification of the muon anomalous magnetic moment $g-2$ due to electron loop insert in higher order. Since the virtual electron pair is in close proximity to the muon, it experiences strong field phenomena. We show that the current theory-experiment muon $g-2$ discrepancy could originate in the (virtual) electron mass  \textit{non-perturbative}  modification by the strong muon EM field.  The magnitude of the electron mass modification can be also assessed via enhancement of $e^+e^-$-pair production in strong fields.
\end{abstract}

\pacs{11.15.Tk,12.20.-m,13.40.-f,13.40.Dk}

\maketitle

\section{Introduction}
The Higgs minimal coupling mass generating mechanism of heavy, beyond GeV mass scale, elementary particles is well established. The origin of the lightest standard model (SM) electron mass is not expected to be resolved experimentally in the near future: the LHC $pp$-collider is capable of constraining the minimal coupling to factor $\sim100$ above the predicted SM value, and next generation $e^+e^-$-colliders are limited to factor $\sim10$ above the required sensitivity~\cite{Altmannshofer:2015qra}. However, given the small value of electron mass a significant electron mass contribution from electromagnetic (EM) self-energy in the realm of QED can be expected.

We propose here a complementary  probe of electron mass, the mass modification by a strong external field. This is based on the observation that a negligible in magnitude beyond SM (BSM) component is irrelevant, while the Higgs mass component can respond to external EM field strengths of electro-weak natural strength, inaccessible today. However, the electromagnetic (EM) mass component of the electron is susceptible to modification by an applied strong external field, measured in electron mass $m_e$ natural units characterized by the Schwinger (EHS) field,
\begin{align} 
\label{Ecrit}
\mE_{\mathrm{EHS}}=\frac{m_e^2(0)}{e}=1.323\times 10^{18}\,\mathrm{V/m}
\;. 
\end{align}

The electron mass response to external EM fields has  been studied before~\cite{Ritus:1970,Shovkovy:2012zn,Ferrer:2015wca}. We revisit the topic with the aim to better understand the relation between the two dominant contributions to electron mass, the EM and Higgs portions.  The EM mass in external fields   can be studied in QED either as EM-self-energy or in the EM-energy-momentum tensor analysis: We  show this  in Sect.~\ref{selfconsistent}, where we also propose an effective method  allowing exploration of field-dependent mass in the strong field nonperturbative regime.  We show  how the  nonlinear mixing between fields suppresses the total field to an amount that can be smaller than their linear superposed sum. For this reason the electron's EM field mass portion diminishes in an externally applied electric field, an effect we call in the following mass melting.
 
In Sect.~\ref{similarity}  using a specific limiting EM field strength  model  we explore the supercritical field regime~\cite{Rafelski:1972fi,BialynickiBirula:1984tx}.  The model is tailored to match closely to the QED effective Euler-Heisenberg-Schwinger (EHS) action~\cite{Heisenberg:1935qt,Weisskopf:1996bu,Schwinger:1951nm,Dunne:2004nc} for quasi-constant fields up to the EHS field strength, and contains an adjustable parameter that allows for probing of the relation between EM and non-EM components of electron mass. In Sect.~\ref{ModelMM} we compute the model mass melting effect in  supercritical fields, and in Sect.~\ref{2loop} show that  in subcritical fields this effective modification has the right structure (powers in $\alpha$) and order of magnitude to be consistent with evaluation of perturbative self-energy corrections in QED.

Most high precision QED experiments probe fields below the EHS critical field, in which the mass modification can be explored using perturbative QED. Fields beyond the EHS strength appear in the muon $g-2$: The anomaly in the  muon magnetic moment is sensitive to the melting of the (virtual) electron mass entering the vacuum polarization experienced by the virtual photon, induced by the muon-sourced EM field. Due to the small effective size of the muon,  as a quantum wave localized at its Compton wave length scale, the virtual electrons entering the muon $g-2$ consideration experience much stronger fields than that in the electron $g-2$ case. We show in Sect.~\ref{PrecisionQEDeffects} via computation of the induced vacuum polarization displacement current charge density that the polarized virtual electron pairs lie close enough to the muon to experience strong field phenomena. We argue that therefore a perturbative QED evaluation of $g-2$ is unreliable.   We show that our model mass melting of the electron by the field of the muon  is capable to explain the observed theory-experiment $g-2$ discrepancy~\cite{Jegerlehner:2009ry}. 

Another experimental process highly sensitive to the value of the electron mass is the QED vacuum decay into electron-positron pairs, see Sect.~\ref{PairProduction}. For uniform homogenous EM fields, pair production in strong fields is inherent in the EHS effective action obtained for quasi-constant fields. For inhomogeneous fields the non-perturbative process of pair production has been explored in the context of heavy ion collisions~\cite{Greiner:1985ce}. The localization of strong field phenomena has facilitated consideration of the local vacuum structure and instability~\cite{Rafelski:1974rh}. Pair production in the EHS action seen in perspective of the development of novel ultra-intense-pulsed-laser technologies~\cite{Mourou:2006zz} has been spurring a renaissance of strong field physics~\cite{Gies:2008wv,Dunne:2008kc,DiPiazza:2011tq,Hegelich:2014tda}. We close this paper with an outline of future research opportunities.

\section{Mass melting arising in nonlinear electromagnetism}
\label{selfconsistent}

\subsection{Effective Scalar Potential}
\label{scalarform}

The mass response to external EM fields is recognized exploring the position of the electron propagator pole, see e.g.~\cite{Itzykson:1980rh}. The field-dependence of electron self-energy~\cite{Ritus:1970}  allows us to  write
\begin{align}
\label{phiform}
m_e(a,b)\equiv m_e  +\phi(m_e;a,b)
\;.
\end{align}
Only in the limit of quasi-constant fields  can $\phi$ in Eq.\,(\ref{phiform}) be expressed as a function of both Lorentz and gauge invariant quantities. Here $m_e(0)\equiv m_e$ is the physical electron mass; the invariants  
\begin{align}
\label{abform}
a^2-b^2= 2S=\mE^2-\mB^2\;\;,\;\;\;\;\;
a^2 b^2= P^2=(\mE\cdot\mB)^2
\;,
\end{align}
are appearing in the eigenvalues of the EM field tensor. The scalar  $\phi$ itself now enters the Dirac equation since
\begin{align} 
\label{diracphi}
\Big[\gamma\cdot \Pi- m_e(a,b) ) \Big]\psi=0
\;,
\end{align}
in view of Eq.\,(\ref{phiform}).

The scalar potential $\phi$ has been studied applying perturbative QED methods: The leading self-energy diagram  consisting of the dressed (by external fields) electron propagator along a virtual photon loop~\cite{Ritus:1970,Ferrer:2015wca}; and magnetic field driven mass (mass catalysis~\cite{Shovkovy:2012zn}, and references therein). The perturbative approach is assumed to be valid for weak fields well below the EHS critical field scale. The supercritical regime requires extending the computation to a nonperturbative summation in self-energy diagrams. This includes a self-consistency requirement  in Eq.\,(\ref{phiform}): On RHS  the mass scale accompanying $a,b$ dependence can be interpreted as being  affected by the EM response. 

To explore this mass response in the nonperturbative regime we propose to consider a  model formulation of $m_{\mathrm{EM}}$, the EM field mass portion of electron mass $m_e$. How a finite $m_{\mathrm{EM}}$ with a stable EM stress configuration arises remains an open question today~\cite{Rafelski:1972fi,Schwinger:1983nt, BialynickiBirula:1993ce}. Electron mass cannot be entirely EM in origin due to lepton mass differences: to resolve the EM portion is not possible within perturbative QED, but may arise in a self-consistent and nonperturbative approach, see~\cite{Wilczek:2012sb}. 
By EM field mass we refer to the energy contained within the EM field of a particle: this quantity is strongly dependent on the particle charge distribution -- in QED the \lq electron size\rq\ is governed by the Compton wavelength, while in classical EM theory the $\alpha^{-1}=137 $ times smaller so-called classical electron radius is the appropriate scale.

\subsection{EM field mass}

We evaluate $m_e(a,b)$ by integrating EM field mass density. The Lorentz invariant field mass density $U$ obtained from the field 4-momentum density $P^\nu$, see for example Eq.\,(28.11) in~\cite{Rafelski:2017hyt}:
\begin{align} 
\label{Utmunu}
P^\nu=&\;u_\mu T^{\mu\nu}\;, \\ 
U\equiv \sqrt{P^\nu P_\nu} =&\;\sqrt{u_\mu T^{\mu\nu}T_{\nu\alpha}u^\alpha}\;, \\
 U =&\;
\sqrt{(T^{00})^2+(T^{0k})^2}
\;,
\end{align}
where $T$ is the EM energy-momentum tensor and $u$ is the relative 4-velocity between observer and the source of the field. The last relation follows in the comoving (relative rest) reference frame of the field source $u=(1,\vec 0)$\;.

$T$ for any nonlinear effective EM Lagrangian $\mL$ is obtained by varying $\mL$ with respect to the metric $g_{\mu\nu}$, resulting in~\cite{Labun:2008qq}:
\begin{align} 
\label{tmunu}
T_{\mu\nu}=\frac{\pt\mL}{\pt S}T_{\mu\nu}^{\mathrm{M}}-g_{\mu\nu}
\Big(\mL-S\frac{\pt\mL}{\pt S}-P\frac{\pt\mL}{\pt P}\Big)
\;,
\end{align}
where $T^{\mathrm{M}}$ is the Maxwell energy momentum tensor
\begin{eqnarray}
T^{\mathrm{M}}_{\mu\nu}(x)=
T_{\mu\nu}(x)\Big|_{\mL=\mL^{\mathrm{M}}}=
F_{\mu\alpha}F^\alpha_{\;\nu}-\frac{g_{\mu\nu}}4F_{\beta\alpha}F^{\alpha\beta}
\;.
\end{eqnarray}
 $g$ is the space-time (here Minkowski) metric, $F$ is the EM field tensor, and the Maxwell Lagrangian
\begin{eqnarray}
\mL^{\mathrm{M}}=\frac12(\mE^2-\mB^2)
\;.
\end{eqnarray}
For a pure electric field, Eq.\,(\ref{tmunu}) may be written as 
\begin{align} 
\label{tmunu2}
T_{\mu\nu}=\frac{\mD}{\mE}\Big(T_{\mu\nu}^{\mathrm{M}}+g_{\mu\nu}\frac{\mE^2}2\Big)-g_{\mu\nu}\mL
\;,
\end{align}
where we introduced the displacement field
\begin{align} 
\mD=\frac{\pt\mL}{\pt \mE}
\;.
\end{align}
Plugging Eq.\,(\ref{tmunu2}) into Eq.\,(\ref{Utmunu}) produces
\begin{align} 
\label{Uform}
U=&\;
\sqrt{\Big(\frac{\mD}{\mE}\Big(T_{00}^{\mathrm{M}}+g_{00}\frac{\mE^2}2\Big)-g_{00}\mL\Big)^2}
\nonumber \\
=&\;
\sqrt{\Big(\mE\cdot\mD-\mL\Big)^2}
\;.
\end{align}
The same result is obtained by performing a Legendre transform of $\mL$~\cite{BialynickiBirula:1984tx}. Note that $U$ and $\mE$ are to be evaluated as functions of $\mD$ which is sourced by some applied charge distribution.

Consider the electric field $\mE_{\mathrm{p}}$ sourced by the charge of the electron (we postpone for now the magnetic field sources by the magnetic dipole moment), in the presence of an external constant and homogeneous electrical field $\mE_{\mathrm{ex}}$. First we write the EM field mass for the electron's electric component in the absence of external fields
\begin{align} 
\label{mzero}
m_{\mathrm{EM}}(\mE_{\mathrm{ex}}=0)=&\;
\int d^3r\,U (\mD_{\mathrm{p}})
\;,
\end{align}
where for a point electron (or, outside of charge distribution of an electron) we have 
\begin{align} 
\mD_{\mathrm{p}}=\frac{\pt\mL}{\pt\mE_{\mathrm{p}}}=\hat r\frac{e}{4\pi|\vec r\,|^2}
\;.
\end{align}
To include an external field, a closer look at the superposition principle is required. Only for Maxwell action may $\mE$ be written as a linear superposition of $\mE_{\mathrm{p}}$ and $\mE_{\mathrm{ex}}$. In any nonlinear theory, the superposition of electric fields is violated, and only the displacement fields generated by inhomogeneous Maxwell equations may be superposed. We use displacement fields to distinguish the external and particle field in the rest frame: 
\begin{align} 
\label{Dsum}
\mD=\mD_{\mathrm{p}}+\mD_{\mathrm{ex}}
=\hat r\frac{e}{4\pi|\vec r\,|^2}+\mD_{\mathrm{ex}}
\;.
\end{align}
The superposed displacement fields enter the expression for mass density (Eq.\,(\ref{Uform})), from which we subtract the contribution from the external field alone:
\begin{align} 
\label{Uex}
\tilde U(\mD)=U(\mD_{\mathrm{p}}+\mD_{\mathrm{ex}})-U(\mD_{\mathrm{ex}})
\;.
\end{align}
Eq.\,(\ref{Uex}) is integrated to obtain the external field-dependent mass:
\begin{align} 
\label{mex}
m_{\mathrm{EM}}(\mE_{\mathrm{ex}})=\int d^3r\,\tilde U(\mD)
\;.
\end{align}
This expression for EM field mass is applicable to both linear and nonlinear actions. 

Only in the linear Maxwell theory is the mass unaffected by the presence of external fields:  the Maxwell Lagrangian
\begin{align} 
\mL^{\mathrm{M}}=\mE^2/2
\;,
\end{align}
gives superposable fields
\begin{align}
\label{superposed}
\mE=\mE_{\mathrm{p}}+\mE_{\mathrm{ex}}
=\mD=\mD_{\mathrm{p}}+\mD_{\mathrm{ex}}
\;,
\end{align}
(only in Gauss-type units can we set $\mE=\mD$, in SI units there is a further vacuum dielectric constant factor). The mass density becomes
\begin{align} 
\tilde U(\mD)=\frac{\mD_{\mathrm{p}}^2}2+\mD_{\mathrm{p}}\cdot\mD_{\mathrm{ex}}
\;.
\end{align}
Evaluating the field mass according to Eq.\,(\ref{mex})
\begin{align} 
m_{\mathrm{EM}}(\mE_{\mathrm{ex}})\Big|_{\mL=\mL^{\mathrm{M}}}
=&\,
\int d^3r\, \Big(\frac{\mD_{\mathrm{p}}^2}2+\mD_{\mathrm{p}}\cdot\mD_{\mathrm{ex}}\Big)
\nonumber \\
=&\,
\int d^3r\,\frac{\mD_{\mathrm{p}}^2}2
=\;
m_{\mathrm{EM}}(\mE_{\mathrm{ex}}=0)
\;,
\end{align}
where the mixing term integrates to zero for any radial Coulomb field in presence of a constant and homogeneous external field. 
Our intuitive chain of argument using classical fields shows clearly that only when the superposition principle of fields holds can one expect to describe the particle interaction with electron mass unchanged by the external field. 

Upon field quantization, that is in QED, the situation becomes more complex since any interacting quantum field theory is intrinsically a nonlinear theory. However, to lowest order the Maxwell Lagrangian would then include the vacuum polarization contribution, described diagrammatically by an electron loop coupled to two photon lines (of order $\mE^2$). Such a contribution though being nonlocal, is still linear and thus does not introduce external field corrections to the EM field mass.

The correction we anticipate appears at higher order. For the case of (quasi-)constant fields, an exact result was obtained by Euler and Heisenberg, and illuminated by Schwinger~\cite{Heisenberg:1935qt,Weisskopf:1996bu,Schwinger:1951nm}. This EHS field-dependent action is giving the vacuum state the properties of a nonlinear dielectric and introduce light-light scattering diagrams, beginning with order $\mE^4$. We have as action to lowest order the Maxwell term complemented by nonlinear effective QED term:
\begin{align}
\label{leffgeneral}
\mL_{\mathrm{EHS}}^1=\frac{\mE^2}2+\frac{\mE^4}{\mE_{\mathrm{cr}}^2}+\ldots
\;,
\end{align}
here expanded to leading fourth order contribution. $\mE_{\mathrm{cr}}$ is a `critical' field scale we recognize in magnitude in quantitative consideration of QED effects. 

The new nonlinear term is mixing the particle and the external field in a more complex fashion due to violation of the EM field superposition principle. In order to use the superposition principle for the displacement fields we evaluate the relation between field and displacement field, and its inversion, in the weak field limit
\begin{align}
\label{suppress}
\mD=\mE+4\frac{\mE^3}{\mE_{\mathrm{cr}}^2}+\ldots
\ \to\ 
\mE=\mD-4\frac{\mD^3}{\mE_{\mathrm{cr}}^2}+\ldots
\;,
\end{align}
see~\cite{Gitman:2012qx}. The electric field is suppressed by the light-light scattering response; hence the negative contribution to $\mE$ on the RHS of Eq.\,(\ref{suppress}). The total electric field is thus smaller than the superposed fields encountered in the linear theory in Eq.\,(\ref{superposed}), reducing the EM field mass density:
\begin{align}
U(\mD)=\frac{\mD^2}2-\frac{\mD^4}{\mE_{\mathrm{cr}}^2}-\ldots
\;.
\end{align}
Writing $\mD=\mD_{\mathrm{p}}+\mD_{\mathrm{ex}}$ explicitly and subtracting the external field contribution far from the Coulomb field source,
\begin{align}
\label{Ueffgeneral}
&\tilde U(\mD)=\frac{\mD_{\mathrm{p}}^2}2+\mD_{\mathrm{p}}\cdot\mD_{\mathrm{ex}}
-\frac1{\mE_{\mathrm{cr}}^2}\Big\{
\mD_{\mathrm{p}}^4+2\mD_{\mathrm{p}}^2\mD_{\mathrm{ex}}^2
\nonumber \\
&
+4\mD_{\mathrm{p}}^2\mD_{\mathrm{p}}\cdot\mD_{\mathrm{ex}}+4\mD_{\mathrm{ex}}^2\mD_{\mathrm{p}}\cdot\mD_{\mathrm{ex}}
+4(\mD_{\mathrm{p}}\cdot\mD_{\mathrm{ex}})^2
\Big\}-\ldots
\;.
\end{align}
Plugging Eq.\,(\ref{Ueffgeneral}) into Eq.\,(\ref{mex}), odd powers $\mD_{\mathrm{p}}\cdot\mD_{\mathrm{ex}}$ integrate to zero and we obtain, to order $\mD_{\mathrm{p}}^2$,
\begin{align} 
\label{meltgeneral}
m_{\mathrm{EM}}(\mE_{\mathrm{ex}})
=&\,
\int d^3r\, \Big(
\frac{\mD_{\mathrm{p}}^2}2
-\frac{2\mD_{\mathrm{p}}^2\mD_{\mathrm{ex}}^2
+4(\mD_{\mathrm{p}}\cdot\mD_{\mathrm{ex}})^2}{\mE_{\mathrm{cr}}^2}
\Big)
\;.
\end{align}
In Eq.\,(\ref{meltgeneral}) the leading correction to the EM field mass density is quadratic in external fields, and due to its negative sign, mass decreases (melts) in an external field.  We now turn to a nonperturbative formulation, essential to a quantitative study of EM field mass.

\section{Nonlinear effective action}
\label{similarity}

We study mass melting in context of the following EM actions:
\subsubsection{EHS effective action:}
In QED, in the quasi-constant (local) field approximation, the EHS effective action arises~\cite{Dunne:2004nc}
\begin{align} 
\label{EHS}
\mL_{\mathrm{EHS}}=&\;\mL^{\mathrm{M}}
-\frac1{8\pi^2}\int_{0}^\infty\frac{ds}{s^{3-\delta}}e^{-m_e^2(0)s}
\nonumber \\
&
\times
\Big(\frac{e^2abs^2 \cot[eas]}{\tanh[ebs]}-1\Big)
\;,
\end{align}
depending implicitly on the Schwinger (EHS) field, Eq.\,(\ref{Ecrit}). The argument `0' seen above for the electron mass reminds that the result was obtained without allowing for a dependence of electron mass on the external applied field.

The field invariants $a$ and $b$ in Eq.\,(\ref{EHS}) are given by Eq.\,(\ref{abform}), and the pre-factor $1/8\pi^2$ in Eq.\,(\ref{EHS}) follows units used by Schwinger in which $\alpha=e^2/4\pi$, see Ref.~\cite{Dunne:2004nc}. The function in Eq.\,(\ref{EHS}) has been subtracted to remove the zero-point energy, and the logarithmically divergent contribution to be absorbed by charge renormalization is regularized by infinitesimal $\delta$. 

In the perturbative regime, weak field expansion ($\mE_{\mathrm{ex}},\mB_{\mathrm{ex}}\ll\mE_{\mathrm{EHS}}$) produces the light-light scattering contribution
\begin{align} 
\label{EHSRbis}
\mL_{\mathrm{EHS}}&=\mL^{\mathrm{M}}+\frac{2\alpha}{45\pi\mE_{\mathrm{EHS}}^2}
\Big\{
S^2+\frac 7 4 P^2+\mathcal{O}(S^3)
\Big\}
\;.
\end{align}

Understanding of the magnitude of the mass melting effect requires  a finite computable EM field mass, see Eq.\,(\ref{mex}), yet the EHS action, without improvements, leads to a divergent result. We demonstrate this divergence in two steps: first considering the divergent electric component of the Maxwell EM field mass and later EHS. For the Maxwell case we find: 
\begin{align} 
m_{\mathrm{EM}}(0)\Big|_{\mL=\mL^{\mathrm{M}}}
=&\,\int d^3r\, \Big( \mE_{\mathrm{p}}\cdot\mD_{\mathrm{p}}- \mL^{\mathrm{M}}(\mE_{\mathrm{p}})\Big)
\nonumber \\
=&\,
\int d^3r\,\frac{\mE_{\mathrm{p}}\cdot\mD_{\mathrm{p}}}2
\nonumber \\
=&\,
\frac{e^2}{8\pi}\int dr\, r^2\cdot \frac{1}{r^4}\to \infty
\;.
\end{align}
To remedy the divergent Maxwell expression requires an effective action which suppresses $\mE$, where at the origin $\mE_{\mathrm{p}}\ll\mD_{\mathrm{p}}$, for example such that product $\mE_{\mathrm{p}}\cdot\mD_{\mathrm{p}}\propto 1/r^2$, instead of $1/r^4$ as in the Maxwell case. Now if the original EHS action were applied instead,
\begin{align} 
m_{\mathrm{EM}}(0)\Big|_{\mL=\mL_{\mathrm{EHS}}}
=&\,\int d^3r\, \Big( \mE_{\mathrm{p}}\cdot\mD_{\mathrm{p}}- \mL_{\mathrm{EHS}}(\mE_{\mathrm{p}})\Big)
\;,
\end{align}
where at the origin the strong field limit ($\mE_{\mathrm{p}}\gg\mE_{\mathrm{EHS}}$) gives
\begin{align} 
\label{strongEHS}
\mathrm{Re}[\mL_{\mathrm{EHS}}]&=\frac{\mE_{\mathrm{p}}^2}{2}
\Big(1-\frac{\alpha}{3\pi}\ln[2e\mE_{\mathrm{p}}/m_e^2(0)]\Big)
\;.
\end{align}
Differentiating Eq.\,(\ref{strongEHS}) by $\mE_{\mathrm{p}}$ to obtain $\mD_{\mathrm{p}}(\mE_{\mathrm{p}})$, we find that $\mE_{\mathrm{p}}>\mD_{\mathrm{p}}$, causing the product $\mE_{\mathrm{p}}\cdot\mD_{\mathrm{p}}$ to be even more divergent than in the Maxwell case.

Four here relevant additional corrections to EHS result are known and we believe can influence the divergent behavior of the EM self-fields above: 
\begin{enumerate}
\item 
In the original EHS result, the fermions are non-dynamical since they are integrated out in order to obtain an effective action: self-energy corrections to the mass can only appear in higher order corrections. Since we know that the EHS effective action is nonlinear, we need to account for the possibility that the electron mass is melted in strong fields. This EM field mass correction described in Sect.~\ref{selfconsistent} enters the Dirac equation at the start of derivation of EHS action, Eq.\,(\ref{diracphi}). One can understand this considering such a self-consistent correction applied to the Landau energy levels summed in the Weisskopf EHS evaluation~\cite{Weisskopf:1996bu}. The result thus is:
\begin{align} 
\label{EHSmelt}
\mL_{\mathrm{EHS+melt}}=&\;\mL^{\mathrm{M}}
-\frac1{8\pi^2}\int_{0}^\infty\frac{ds}{s^{3-\delta}}e^{-m_e^2(a,b)s}
\nonumber \\
&
\times
\Big(\frac{e^2abs^2 \cot[eas]}{\tanh[ebs]}-1
\Big)
\;.
\end{align}
Eq.\,(\ref{EHSmelt}) contains a higher order correction to the EHS action via field-dependent mass, hence subscript \lq+melt\rq. Interestingly, the Schwinger field is now an external field-dependent quantity: 
\begin{align} 
\label{ECritmelt}
\mE_{\mathrm{EHS}}^{\mathrm{melt}}=\frac{m_e^2(a_{\mathrm{ex}},b_{\mathrm{ex}})}{e}
\;.
\end{align}
\item Nonlocal corrections to the action are required in order to account for nonlinear mixing between the external fields and the inhomogeneous fields of the particles~\cite{Dunne:2005sx,Dunne:2006st,Kim:2007pm,Kim:2009pg}. 
\item Another effect has been pointed out by Gies and Karbstein~\cite{Gies:2016yaa,Karbstein:2019wmj}: in addition to the well-studied internal photon line corrections to the EHS action~\cite{Ritus:1975cf,Huet:2017ydx}, reducible connecting photon line corrections to the EHS action have  been shown to be nonvanishing. This contribution may also be accounted for in a self-consistent manner. We continue this work in an upcoming paper,  following a suggestion by Weisskopf that the EM fields entering effective action are themselves screened by the vacuum response~\cite{Weisskopf:1996bu}.

\item The modification to effective action due to QED induced anomalous magnetic  moment has recently attracted attention and we refrain from in depth discussion here. We note that in addition to mass, the $g-2$ contribution becomes a field-dependent quantity~\cite{Ferrer:2015wca} which must enter into EHS action in a self consistent manner. For incorporation of $g-2$ into EHS action see~\cite{OConnell:1968spc,Dittrich:1977ee,Kruglov:2001dp,Labun:2012jf}, recently shown by the authors to have a significant influence on pair production in strong magnetically dominated fields~\cite{Evans:2018kor}. 
\end{enumerate}

\subsubsection{Generalized Born-Infeld model:}
Practical use of EHS action considering the above described self-consistency extensions is difficult as considerable improvement of our understanding of the  (electron) mass response to external fields is required. Therefore we explore a model of the  Born-Infeld  (BI)  type created expressly  to describe EM-mass of the electron. We recall that BI model also introduces a limit to the achievable field strength.  We do not consider the model as an extension to QED, which has in large part been constrained~\cite{Davila:2013wba,Rebhan:2017zdx,Ellis:2017edi}. Instead we fine-tune it so that it will allow to make predictions in lieu of the EHS effective action  accounting perhaps for the required EHS improvements we discussed.

We fine-tune a BI-like model introduced in Ref.~\cite{Rafelski:1972fi} 
\begin{align} 
\label{LeffForm}
\mL_{\mathrm{eff}}=&\;-\frac{\mE_{\mathrm{cr}}^2(n)}{2n}
\bigg(
\Big(1-\frac{a^2-b^2}{\mE_{\mathrm{cr}}^2(n)}-7\frac{a^2b^2}{\mE_{\mathrm{cr}}^4(n)}
\Big)^n-1\bigg)
\;.
\end{align}
Choosing $n=1/2$ and $\mE_{\mathrm{cr}}=89.72\mE_{\mathrm{EHS}}$ we obtain the original BI result~\cite{Born:1934}. $\mE_{\mathrm{cr}}$ provides a limit on the maximum strength that an electric field may reach, and we inserted the $a^2b^2$ coefficient so that the model more closely tracks in its functional form the  EHS action in presence of both electric and magnetic fields. 

\begin{figure}[ht]
\centering
\includegraphics[width=.9\columnwidth]{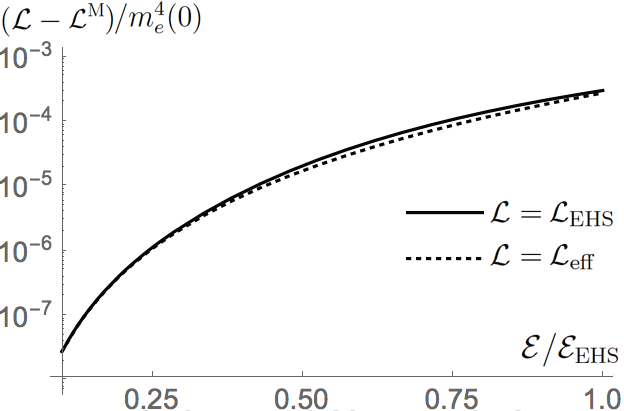}
\caption{\label{similarL} Nonlinear contribution of $\mL_{\mathrm{eff}}$ field for all $-1/2\leq n\leq1/2$ alongside the real part of $\mL_{\mathrm{EHS}}$ ($\mB=0$). Choice of $n$ does not affect $\mL_{\mathrm{eff}}$ until stronger EM fields.} 
\end{figure}

We further choose the value of the critical field such that an expansion of Eq.\,(\ref{LeffForm}) generates exactly the EHS light-light scattering result, Eq.\,(\ref{EHSRbis})  
\begin{align} 
\label{Ecrmatch}
\mE_{\mathrm{cr}}(n)=\mE_{\mathrm{EHS}}\sqrt{\frac{45\pi(1-n)}{2\alpha}}
\;. 
\end{align}
This also assures that $\mL_{\mathrm{eff}}$ and the original $\mL_{\mathrm{EHS}}$ agree in shape up to EHS field $\mE_{\mathrm{EHS}}$, see figure~\ref{similarL}. Because the action  Eq.\,(\ref{LeffForm}) has two free parameters, different choices of $n\leq1/2$ in Eq.\,(\ref{LeffForm}) allow  to choose  the magnitude of  finite EM field mass, as opposed to the original BI model where all electron mass was attributed to the EM field. A finite EM-mass arises in the range $n\leq1/2$ since the product $\mE\cdot\mD\propto 1/r^2$ at the origin. 

%
%
%

\section{Evaluation of effective mass}

\subsection{Model computation}
\label{ModelMM}

\begin{figure}[ht]
\centering
\includegraphics[width=.95\columnwidth]{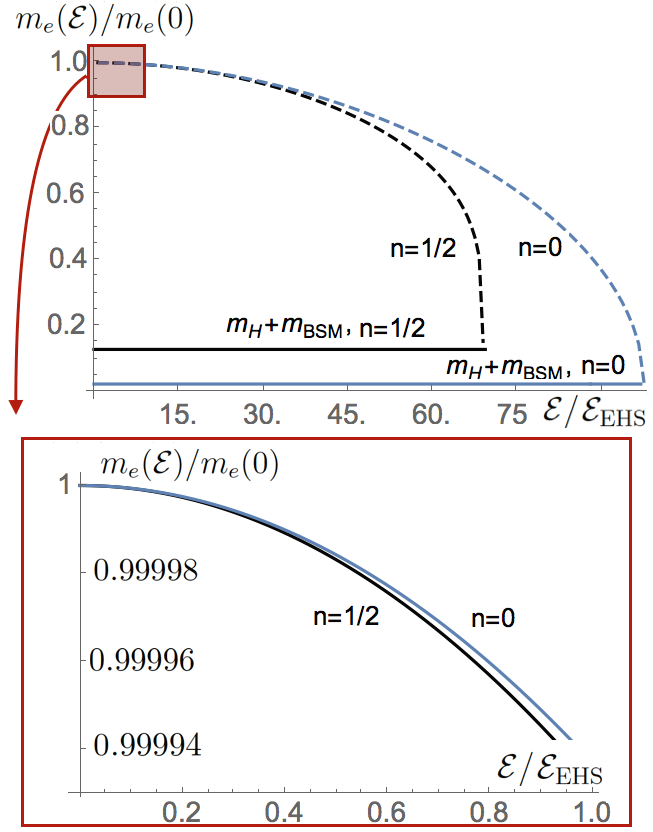}
\caption{\label{Fig1b} Top: dashed lines plot mass-melting models with parameters $n=1/2$ and $n=0$, solid lines for the non-EM mass $m_\mathrm{H}+m_\mathrm{BSM}$. Bottom: zoom-in detail of mass behavior below the EHS field.}
\end{figure}

We compute the electric component of EM field mass via the displacement field, differentiating the model for effective action, Eq.\,(\ref{LeffForm}):
\begin{align} 
\mD=\mD_{\mathrm{p}}+\mD_{\mathrm{ex}}=\frac{\pt\mL_{\mathrm{eff}}}{\pt\mE}
=\mE\Big(1-\frac{\mE^2}{\mE_{\mathrm{cr}}^2(n)}\Big)^{n-1}
\;.
\end{align}
$\mD$ is numerically inverted to obtain an expression for electric field $\mE(\mD)=\mE(\mD_{\mathrm{p}}+\mD_{\mathrm{ex}})$.
We separately define the externally applied electric field $\mE_{\mathrm{ex}}$ as the limit far from the particle:
\begin{align} 
\label{DBI2}
\lim_{|\vec r\,|\to\infty}\mD=\mD_{\mathrm{ex}}
=\mE_{\mathrm{ex}}\Big(1-\frac{\mE_{\mathrm{ex}}^2}{\mE_{\mathrm{cr}}^2(n)}\Big)^{n-1}
\;,
\end{align}
numerically inverted to obtain $\mE_{\mathrm{ex}}(\mD_{\mathrm{ex}})$. $\mE$ and $\mE_{\mathrm{ex}}$ are then plugged into the mass density given by Eqs.\,(\ref{Uex}) and (\ref{mex}):
\begin{align} 
\label{mex2}
m_{\mathrm{EM}}(\mE_{\mathrm{ex}})=&\,
\int d^3r\, 
\Big(U(\mD)-U(\mD_{\mathrm{ex}})\Big)
 \\ \nonumber
=&\,
\int d^3r\, \Big\{
\Big(\mE(\mD)\cdot\mD-\mL_{\mathrm{eff}}(\mE(\mD))\Big)
 \\ \nonumber
&\,
-\Big(\mE_{\mathrm{ex}}(\mD_{\mathrm{ex}})\cdot\mD_{\mathrm{ex}}-\mL_{\mathrm{eff}}(\mE_{\mathrm{ex}}(\mD_{\mathrm{ex}}))\Big)
\Big\}
\;.
\end{align}
Eq.\,(\ref{mex2}) is computed numerically and plotted in figure~\ref{Fig1b} for the examples of $n=1/2$ and $n=0$. 
%
%
%

The plots in figure~\ref{Fig1b} are a model example. We see the significant non-perturbative behavior and in the zoom-insert at bottom the domain which explains why one can often argue that mass-melting is a minor effect. Total melting of the EM field mass occurs at the critical fields $\mE_{\mathrm{cr}}(n)\gg\mE_{\mathrm{EHS}}$: $\mE_{\mathrm{p}}$ is suppressed as the externally applied field $\mE_{\mathrm{ex}}$ approaches $\mE_{\mathrm{cr}}(n)$, due to the violation of linear superposition (see Sect.~\ref{selfconsistent}). That is, when the total field cannot exceed $\mE_{\mathrm{cr}}(n)$ and the external field approaches this limit, there is no room left for the particle field. At this point only the (model $n$-dependent) non-EM mass (Higgs+BSM) components remain, flat solid lines in figure~\ref{Fig1b}.


\subsection{Perturbative self-consistent corrections}
\label{2loop}

The leading EM self-energy correction is known via the perturbative QED computation of the internal photon line correction to EHS action. To ensure that the effective form studied here is consistent, we compare the two and show that the model effect is of the right order.

To understand the weak field behavior of the mass melting effect we expand the integrand in Eq.\,(\ref{mex2}) in powers of $\mE_{\mathrm{ex}}$ (from hereon we drop subscript \lq ex\rq) and integrate to obtain the light-light scattering contribution. This is the quadratic in EM field mass modification: 
\begin{align}
\label{mBIpert}
m_{\mathrm{EM}}(\mE)\Big|_{n=1/2}
=&\;
m_{\mathrm{EM}}(0)\Big(1-\frac{\alpha}{30\pi}\frac{\mE^2}{\mE_{\mathrm{EHS}}^2}-\ldots\Big)
\;,
\nonumber \\
m_{\mathrm{EM}}(\mE)\Big|_{n=0}
=&\;
m_{\mathrm{EM}}(0)\Big(1-\frac{\alpha}{36\pi}\frac{\mE^2}{\mE_{\mathrm{EHS}}^2}-\ldots\Big)
\;.
\end{align}
The EM field mass $m_{\mathrm{EM}}(0)$ makes up the following portions of total electron mass for $n=1/2$ and $n=0$: $0.8703m_e(0)$ and $0.9754m_e(0)$, figure~\ref{Fig1b} top. Counting powers in $\alpha$, the mass modification of order $\alpha\mE^2/\mE_{\mathrm{EHS}}^2$ corresponds to the leading perturbative QED treatment, described diagrammatically by the electron propagator connected to a virtual photon loop that encloses two external photon lines~\cite{Ritus:1970}.

We track how the effective action is modified by mass melting in a self-consistent manner. Eq.\,(\ref{mBIpert}) is input into the mass entering EHS action for a pure electric field:
\begin{align} 
\mL_{\mathrm{EHS+melt}}=&\;\mL^{\mathrm{M}}
-\frac1{8\pi^2}\int_{0}^\infty\frac{ds}{s^{3-\delta}}e^{-m_e^2(\mE)s}
\nonumber \\
&\qquad\qquad\quad
\times
\Big(e\mE s \cot[e\mE s]-1\Big)
\;.
\end{align}
Writing explicitly the bare charge in the Maxwell contribution to demonstrate renormalization procedure and expanding to order $\mE^4$,
\begin{align} 
&\mL_{\mathrm{EHS+melt}}
\nonumber \\
=&\;\frac{e^2\mE^2}{2e_0^2}
+\frac1{8\pi^2}\int_{0}^\infty\frac{ds}{s^{3-\delta}}e^{-m_e^2(\mE)s}
\Big(\frac{(e\mE s)^2}3+\frac{(e\mE s)^4}{45}\Big)
\nonumber \\
=&\;
\frac{e^2\mE^2}{2e_0^2}
\Big\{1+\frac{e_0^2}{12\pi^2}\Big(
\delta^{-1}-\gamma_E-\ln[m_e^2(0)]
-\ln\Big[\frac{m_e^2(\mE)}{m_e^2(0)}\Big]
\Big)\Big\}
\nonumber \\
&\;+\frac{\alpha\mE^4}{90\pi\mE_{\mathrm{EHS}}^2}
\;.
\end{align}
We apply renormalized charge
\begin{align} 
\frac1{e^2}=\frac{1}{e_0^2}
\Big(1+\frac{e_0^2}{12\pi^2}\big(\delta^{-1}-\gamma_E-\ln[m_e^2(0)]\big)\Big)
\;,
\end{align}
and use effective mass model Eq.\,(\ref{mBIpert}) to write the remaining finite logarithmic expression to order $\mE^2$,
\begin{align} 
\ln\Big[\frac{m_e^2(\mE)}{m_e^2(0)}\Big]_{n=1/2}=&\;\frac{0.8703\,\alpha}{30\pi}\frac{\mE^2}{\mE_{\mathrm{EHS}}^2}
\;,
\nonumber \\
\ln\Big[\frac{m_e^2(\mE)}{m_e^2(0)}\Big]_{n=0}=&\;\frac{0.9754\,\alpha}{36\pi}\frac{\mE^2}{\mE_{\mathrm{EHS}}^2}
\;.
\end{align}

To order $\mE^4$ the effective action becomes
\begin{align} 
\mL_{\mathrm{EHS+melt}}\Big|_{n=1/2}=&\;\mL^{\mathrm{M}}+\frac{\alpha\mE^4}{90\pi\mE_{\mathrm{EHS}}^2}
\Big(1+\frac{0.8703\,\alpha}{2\pi}\Big)\;,
\nonumber \\ \label{EffMassCor}
\mL_{\mathrm{EHS+melt}}\Big|_{n=0}=&\;\mL^{\mathrm{M}}+\frac{\alpha\mE^4}{90\pi\mE_{\mathrm{EHS}}^2}
\Big(1+\frac{0.9754\cdot5\,\alpha}{12\pi}\Big)
\;,
\end{align}
 aligning in order of $\alpha$ well with the known 2-loop QED correction to EHS action~\cite{Ritus:1975cf,Dunne:2004nc,Dittrich:1985yb}, 
\begin{align} \label{MassCorQED}
\mL_{\mathrm{EHS+2loop}}&=\mL^{\mathrm{M}}+\frac{\alpha\mE^4}{90\pi\mE_{\mathrm{EHS}}^2}
\Big(1+\frac{40\alpha}{9\pi}\Big)
\;.
\end{align}
We see that the EM field effective mass correction Eq.\,(\ref{EffMassCor}) is smaller compared to QED perturbative result Eq.\,(\ref{MassCorQED}).

For higher order in $\alpha$ and $\mE$ effects, the self-consistency requirement means that the corrected \lq EHS+melt\rq\ effective action must again be plugged back into EM field mass computation. Repeating this procedure forces us to consider higher orders in the semi-convergent perturbative EHS series, along with higher order radiative corrections described above in Sect.~\ref{similarity}. Thus a full self-consistent computation is in principle nonperturbative and the radiative corrections have to be built in.

\section{Methods for measuring mass melting model}
\label{MMexp}

\subsection{Muon magnetic moment}
\label{PrecisionQEDeffects}

\begin{figure}[ht]
\centering
\includegraphics[width=.9\columnwidth]{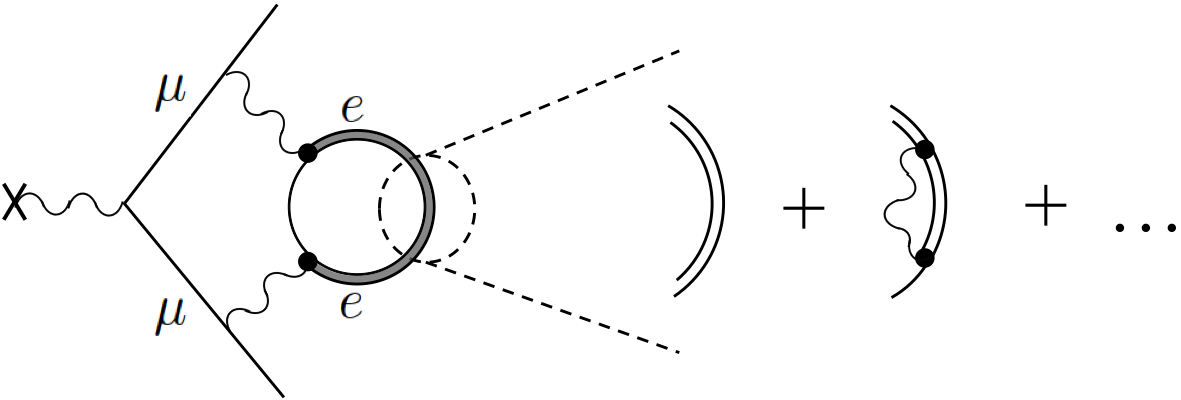} 
\caption{\label{Fig4}Electron vacuum polarization contributions to $\mu$AMM: \lq{x}\rq\ denotes $\mB_{\mathrm{ex}}$. } 
\end{figure}

Turning our attention to precision QED experiments, we consider the muonic $g-2$ anomalous magnetic moment ($\mu$AMM). Mass melting affects the electron mass entering the vacuum polarization contributions to the $\mu$AMM. The muon sees $\mB_{\mathrm{ex}}$, the external magnetic field applied in measuring the $\mu$AMM, while the virtual electron sees both $\mB_{\mathrm{ex}}$ and the electromagnetic fields of the muon $\mE_\mu,\mB_\mu$. Figure~\ref{Fig4} shows the contribution to the $\mu$AMM in which the electron loop is \lq doubly-dressed\rq: zoom-in shows the propagator dressed by the muon's fields (two-line), and summation of its self-energy corrections -- the two-line propagator sums all  photons connecting the electron loop to the muon, and the self-energy correction sums virtual photons that only couple to the electron.

%
%
%

The electron vacuum polarization correction to the $\mu$AMM  is given by Eq.\,(77) in~\cite{Jegerlehner:2009ry}
\begin{align}
\label{deltaa}
\Delta a_\mu(m_e(\mE))\sim\frac{\alpha^2}{\pi^2}\Big(\frac13\ln\Big[\frac{ m_\mu }{m_e(\mE)}\Big]-\frac{25}{36}\Big)
\;.
\end{align}
We have extended this well known expression by allowing the electron mass to be function of the EM field, \textit{i.e.} to melt. The $\ln [m_\mu/m_e]$ term is sensitive in nonperturbative fashion to value of $m_e$ that is subject to mass melting. We did not suggest that the muon mass melts -- we expect that the electron proportionally has more EM field mass than the muon,  and the muon requires much stronger fields (on the order of $m_\mu^2/e$) for noticeable EM field mass modification to occur.

\begin{figure}[ht]
\centering
\includegraphics[width=.9\columnwidth]{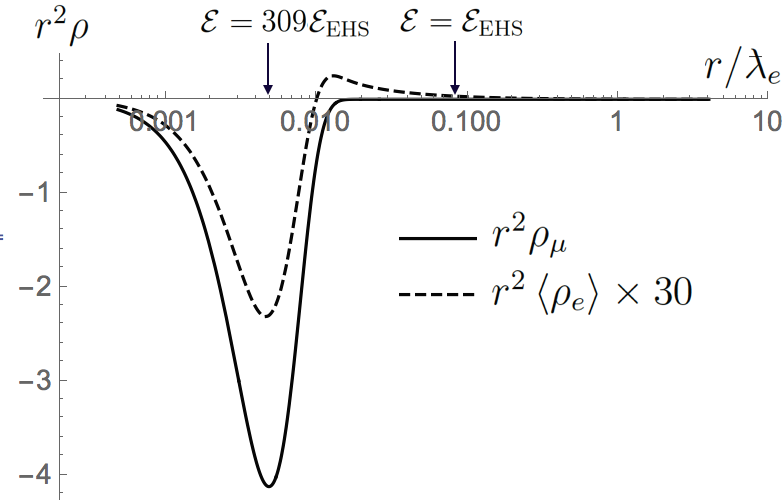}
\caption{\label{Figmu}Solid: muon charge distribution, dashed: induced electron vacuum charge. } 
\end{figure}

\begin{figure}[ht]
\centering
\includegraphics[width=.9\columnwidth]{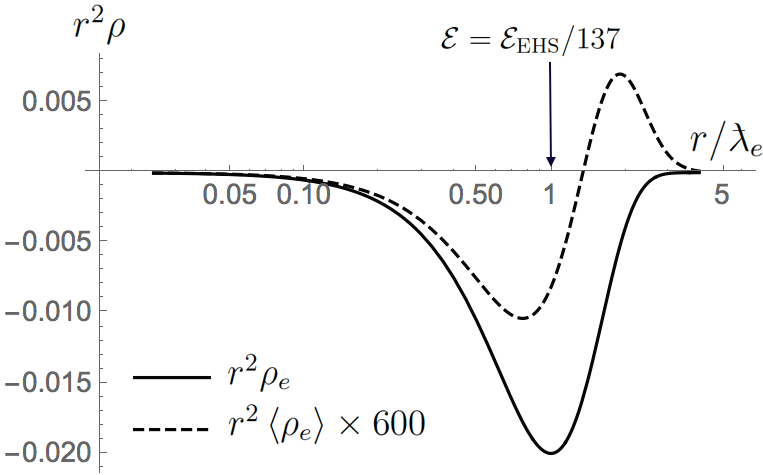}
\caption{\label{Fige}Solid: electron charge distribution, dashed: induced electron vacuum charge. } 
\end{figure}

The electron and its mass is  impacted by the strong muon EM field;  we consider this remark in quantitative manner by estimating the relative  virtual electron pair location with respect to the muon. To illustrate that the virtual electrons cannot resolve the muon to distances smaller than the muon Compton wavelength, we assign a charge distribution $\rho_\mu(r)=2\alpha^{1/2}\exp[-r^2/\lambdabar_\mu^2]\pi^{-1}\lambdabar_\mu^{-3}$. This assumes that the muon is localized to the distance of one muon Compton wavelength, much smaller compared to that of the electron: $\lambdabar_\mu=(m_e/m_\mu)\cdot\lambdabar_e=(1/206)\cdot386\,\mathrm{fm}=1.87\,\mathrm{fm}$. The muon does not  possess such a charge distribution, but this provides an estimate for the field strengths that the virtual \lq quantum sized\rq\ particles experience.
We compute the induced perturbative current derived by Schwinger~\cite{Schwinger:1951nm} and repeat this computation replacing the muon by an electron localized to its Compton wavelength. In figures~\ref{Figmu} and~\ref{Fige} we plot the muon and respective electron ($\rho_e(r)=2\alpha^{1/2}\exp[-r^2/\lambdabar_e^2]\pi^{-1}\lambdabar_e^{-3}$) charge distributions, and their induced (virtual electron)  polarization charge clouds.

The induced charge is much closer to the muon than the electron: we mark the muon Compton wavelength at which $\mE_\mu$ is 309 times the EHS field, and further out the radius at which $\mE_\mu$ is equal to the EHS field. We see that  the muon's induced polarization charge occupies the domain of field strengths in which perturbative treatment of the polarization loop contribution to $g-2$ cannot be trusted. The electron's induced charge lies in the perturbative regime and thus the electron's $g-2$ avoids this difficulty. The importance of nonperturbative AMM computation in the presence of strong Coulomb fields was also noted by Sikora and collaborators~\cite{Sikora:2018zda}. However these authors considered the electron AMM only.

%
%
%

%
%
%

To estimate the order of magnitude of the nonperturbative effect we evaluate as an example the electron mass subject to a field strength of 4 times the EHS field, giving $m_e(\mE) =0.99876\,m_e(0)$ and $m_e(\mE) =0.99897\,m_e(0)$, for $n=1/2$ and $n=0$ respectively. The $\mu$AMM increases by 
\begin{align}
\label{meltAMM}
\Delta a_\mu(m_e(\mE))\Big|_{n=1/2}-\Delta a_\mu(m_e(0))=&\;2.23\cdot10^{-9}
\nonumber \\
\Delta a_\mu(m_e(\mE))\Big|_{n=0}-\Delta a_\mu(m_e(0))=&\;1.86\cdot10^{-9}
\;,
\end{align}
close to the discrepancy between experimental and theoretical values of $2.9\cdot10^{-9}$~\cite{Jegerlehner:2009ry}.

We do not discuss in comparable depth the electron-AMM, as it is less affected than the $\mu$AMM. As noted already, due to the difference in Compton wavelengths, virtual electrons exist much farther from a real electron than from a muon, experiencing weaker mass melting fields: at $\lambdabar_e$ the electric field is only $1/137$ of the EHS field, at which perturbative QED is valid and higher order loop diagrams have been highly constrained~\cite{Schlenvoigt}. The model gives a maximum melting effect of $m_e(\mE)\sim m_e(0)(1-10^{-9})$, modifying the electron $g-2$ on the order of one part per $10^{12}$, two orders of magnitude below the experimental uncertainty~\cite{Hanneke:2010au}.

\subsection{Pair production in strong fields}
\label{PairProduction}

\begin{figure}[ht]
\centering
\includegraphics[width=.8\columnwidth]{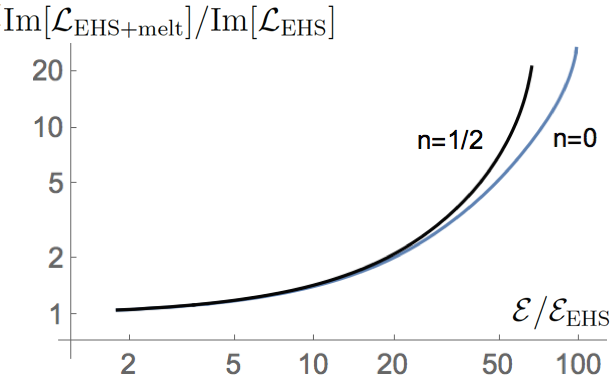}
\caption{\label{Fig3} Enhancement of pair production for effective models with $n=1/2,\;0$, normalized to the original result obtained with constant electron mass.} 
\end{figure}

As ultra short pulse laser fields approach strengths capable of probing vacuum instability, the EHS action has been a focus of much theoretical attention~\cite{Greiner:1985ce,Dittrich:2000zu,Dunne:2004nc}. We explore how the mass melting modifies the imaginary part of EHS action describing the pair production 
\begin{align} 
\label{imEHS}
\mathrm{Im}[\mL_{\mathrm{EHS+melt}}]=&\;
\frac{e^2ab}{8\pi^3}\sum_{n=1}^\infty\frac{\coth[n\pi b/a]e^{-n\pi m_e^2(a,b)/ea}}{n}
\;.
\end{align}
From   Eq.\,(\ref{imEHS}) the rate of pair production and vacuum decay time is obtained~\cite{Labun:2008re}. In figure~\ref{Fig3} the enhancement of pair production according to Eq.\,(\ref{imEHS}) with mass modified by the model computation, Sect.~\ref{ModelMM}, is shown.

%
%
%

We find that the mass melting model does not significantly modify pair production until EM field strengths beyond $\mE_{\mathrm{EHS}}$ are reached, that is when a significant portion of electron mass has melted. While high intensity lasers are unlikely to be suitable for measuring mass melting until they reach the EHS field strength regime, stronger fields capable of probing the effect are generated in heavy ion collisions~\cite{Rumrich:1991xs,Baur:2007fv,Ruffini:2009hg,Rafelski:2016ixr}.

\section{Outlook}
\label{outlook}
 We proposed an effective formulation of the EM field mass  in the strong field regime  not amenable to perturbative QED evaluation. We discuss the possible experimental implications. 

The external field-dependent electron mass can be obtained considering nonlinear EM theory: The violation of superposition in nonlinear EM theories is described in Sect.~\ref{selfconsistent}. The effective EM field mass given in Eq.\,(\ref{Uform}) causes mass modification, Eq.\,(\ref{meltgeneral}). We have explored this effect using  a  model limiting field effective action adjusted to match the EHS light-light scattering, Eq.\,(\ref{LeffForm}). In our model mass melting aligns in magnitude with perturbative QED for weak fields (up to field-gradient corrections), and produces a large degree of melting in the regime of supercritical fields where an exact QED evaluation is not available, figure~\ref{Fig1b}.

We explored observable consequences of the mass melting in the $\mu$AMM. The EM fields sourced by the muon at the location of virtual polarized $e^+e^-$-pairs are in the regime in which nonperturbative strong field QED computation is required, figure~\ref{Figmu}. Applying the mass melting model prediction we found a narrowing of the experimental result discrepancy with the theory Eq.\,(\ref{meltAMM}). We relate to a proposed resolution to the $\mu$AMM discrepancy that involves an external scalar field~\cite{Liu:2016qwd,Davoudiasl:2018fbb},   since the external field-dependent shift in mass can be described in terms of an effective scalar potential. Our nonperturbative result influences the $\mu$AMM in a manner not visible to perturbative QED evaluation. The model prediction has a negligible effect on the electron AMM, due to the weaker EM fields experienced by the virtual electrons. We have also evaluated a mass melting enhancement of pair production in fields relevant to heavy ion collisions, Sect.~\ref{PairProduction}.

A future area of interest is that of high $Z$ atoms, which also probe mass melting of the electron: An electron in the hydrogen-like uranium $1S$-state is likely subject to mass melting since the atomic nucleus with $Z=92$ provides strong  EM fields, on the order of $\mE_{\mathrm{EHS}}$, and there is ample experimental measurement of the $1S$-state binding energy to probe mass melting~\cite{Mohr:1998grz}. We also comment on muonic hydrogen: for a long time there has been a Lamb-shift discrepancy~\cite{Pohl:2010zza}, which in part motivated this work. However this may have been resolved recently~\cite{Bezginov:2019mdi}. Since the muon and proton fields cancel at least in part, we expect the mass melting effect to be smaller than in the case of the muon $g-2$ discrepancy. 

Regarding future theoretical developments, we return to Sect.~\ref{similarity} where we have discussed how we plan to improve the EHS action for quasi constant fields accounting for field-dependent mass and the recent development in reducible diagram summation~\cite{Gies:2016yaa,Karbstein:2019wmj}. Another modification is the nonlocal correction to account for nonlinear mixing between the external fields and the inhomogeneous fields of the particles~\cite{Dunne:2005sx,Dunne:2006st,Kim:2007pm,Kim:2009pg}.  These improvements must be incorporated in a self-consistent manner: the nonlinear effective action, used to compute the external field-dependent mass via our effective formulation, also contains the corrected mass built in. 

Another topic discussed in Sect.~\ref{similarity} concerns how nonlinear EM model theories may be used to model QED effective action. The model action is a monotonically increasing function that matches the known EHS light-light scattering, yet the two actions differ in stronger fields. A questionable sign flip occurs in the real part of EHS action beyond $\mE_{\mathrm{EHS}}$, which causes the EM field mass to be divergent. This sign flip is in the regime where higher order corrections are capable of significantly altering the action. Since mass melting reduces the EHS field~Eq.\,(\ref{ECritmelt}), we expect a self-consistent computation will produce a compounding effect that becomes important in strong fields, possibly producing a further change in sign.

Study of mass melting should also consider the magnetic moment\rq s contribution to EM field mass, which may affect the predicted rate of mass melting. Such a computation is more involved: so far the generalized model limits only the $\mE$ field~\cite{Rafelski:1972fi}, and an extension of the model to limit both $\mE$ and $\mB$ has not yet been invented: the BI-like models are singular upon inclusion of a point-like magnetic dipole. Since the dipole field is important at a smaller radius than the Coulomb field, whether the electric or magnetic mass component is the dominant contribution depends on the nonzero effective size of the electron that arises~\cite{Schwinger:1983nt}. This size lies in between the two physically relevant length scales: the classical electron radius ($2.82$ fm) and the electron Compton wavelength ($386$ fm). 

The work we presented is a tip of an iceberg of many questions that our insight about the effect of EM field nonlinearity presents, inherent in QED. It is possible that there is not a smooth mass melting but a mass discontinuity that will appear in the presence of both electric and magnetic self-mass sources. Our work could also revive effort to conduct numerical study of QED leading to model independent nonperturbative evaluation of EM field mass melting in the presence of strong fields. 

The methods we presented should allow the development of theoretical description of the EM mass component of the electron in presence of strong external fields. A natural consequence of this QED based re-consideration of electron mass  is the differentiation between intrinsic material mass (due to Higgs field)  and the  EM  electron mass, otherwise inaccessible to present day experiment.

\begin{acknowledgments}
The authors would like to express their gratitude to Dr. Tam\'as Bir\'o and Dr. P\'eter L\'evai for their hospitality at the Wigner Research Center for Physics, Budapest, and at the 2019 Balaton Workshop, where in part this work was carried out in the Summer 2019. Johann Rafelski was a Fulbright Fellow during this period. 
\end{acknowledgments}

\end{document}